\documentclass[english,twocolumn,aps,prl, superscriptaddress,showpacs, amsmath]{revtex4-1}

\usepackage[T1]{fontenc}
\usepackage[latin9]{inputenc}
\usepackage{color}
\usepackage{units}
\usepackage{amssymb}
\usepackage{graphicx}
\usepackage{esint}
\usepackage{bm}
\usepackage{natbib}
\usepackage{ulem}
\usepackage[colorlinks=true, citecolor=blue]{hyperref}
\setcitestyle{journalcolor= blue}

\newcommand{\noise}{${<\delta R_{sheet}^2>}/{<R_{sheet}^2>}$ }
\newcommand{\LAOSTO}{LaAlO$_3$/SrTiO$_3$ }
\newcommand{\vg}{$V_g^*$}
\usepackage{graphicx}
\usepackage{dcolumn}
\usepackage{bm}

\usepackage{babel}

\begin{document}

\title{Effect of multiband transport on charge carrier density fluctuations at the \LAOSTO interface}

\author{Gopi Nath Daptary}
\affiliation{Department of Physics, Indian Institute of Science, Bangalore 560012, India}

\author{Pramod Kumar}
\thanks{Present Address: Department of Physics, St.John's College, Agra, Uttar Pradesh 282 002, India}
\affiliation{National Physical Laboratory, New Delhi 110012, India}

\author{Anjana Dogra}
\affiliation{National Physical Laboratory, New Delhi 110012, India}

\author{Aveek Bid}
\email{aveek@iisc.ac.in}
\affiliation{Department of Physics, Indian Institute of Science, Bangalore 560012, India}

\begin{abstract}

Multiband transport in superconductors is interesting both from an academic as well as an application point of view. It has been postulated that interband scattering can significantly affect the carrier dynamics in these materials. In this article we present a detailed study of the electrical transport properties of  the high-mobility two-dimensional electron gas residing at the interface of LaAlO$_3$/SrTiO$_3$, a prototypical multi-band superconductor. We show, through careful measurements of the gate dependence of the magnetoresistance and resistance fluctuations at ultra-low temperatures, that transport in the superconducting regime of this system has contributions from two  bands which host carriers of very different characters.  We identify a gate-voltage tunable Lifshitz transition in the system and show that the resistance fluctuations have strikingly different features on either side of it. At low carrier densities, resistance noise is dominated by number-density fluctuations arising from trapping-detrapping of charge carriers from defects in the underlying SrTiO$_3$ substrate, characteristic of a single-band semiconductor. Above the Lifshitz transition, the noise presumably originates from inter-band scattering. Our work highlights the importance of  inter-band scattering processes in determining the transport properties of low-dimensional systems and projects resistance fluctuation spectroscopy as a viable technique for probing the charge carrier dynamics across a Lifshitz transition.
\end{abstract}

\maketitle


Multiband transport in superconductors is a very active field of research. It is believed that the enhancement of the superconducting transition temperature $T_C$ seen in many superconductors such as MgB$_2$ \cite{iavarone2002two}, H$_3$S \cite{jarlborg2016breakdown} and FeSe \cite{lin2016multiband} is intricately related to a multiband Lifshitz transition. Multiple condensates in these superconducting material can interfere leading to effects not seen in single-band superconductors. It has been postulated that inter-band scattering can significantly affect the carrier dynamics in these materials.  Deconvoluting the contributions of the Cooper pairs residing in different bands to electrical transport properties has proven to be quite difficult. In the case of theoretical investigations, the complication arises partly from the fact that these superconductors cannot be described by the phenomenological Ginzburg-Landau picture - rather by the Usadel equations which are theoretically more intractable. Thus, despite research for over a decade, there exists many open questions regarding the effect of the presence of multiple types of carriers and their interactions on superconductivity. 
 
We approach this problem by studying the carrier dynamics in a well known  two-dimensional superconductor at temperatures slightly above the superconducting $T_C$. Oxide heterostructure as for example, the interface between (001) oriented SrTiO$_3$ and LaAlO$_3$ [hereafter referred to as LaAlO$_3$/SrTiO$_3$] has emerged as one of the most interesting material in condensed matter physics \cite{ohtomo2004high}. Beyond a certain critical thickness of the LaAlO$_3$  layer, a conducting quasi-two dimensional electron gas (q-2DEG) appears at the interface \cite{thiel2006tunable}. Depending on the temperature $T$ and charge carrier density $n$, this q-2DEG can show a broad range of phenomena including superconductivity, ferromagnetism and Rashba spin-orbit coupling \cite{sulpizio2014nanoscale,hwang2012emergent} (SOC). Despite intense research over the past decade \cite{sulpizio2014nanoscale}, the origin of the charge carriers at the interface remains an open question. It is widely believed that the charge carriers appear at the polar/non-polar interface due to electron transfer from the topmost LaAlO$_3$ layer to the $t_{2g}$ levels of $d$-orbitals of Ti at the interface \cite{mannhart2008two}. The three-fold degeneracy of $t_{2g}$ levels found in bulk SrTiO$_3$ is lifted in the case of LaAlO$_3$/SrTiO$_3$ by quantum confinement induced in the z-direction. This causes the $d_{xy}$ orbital to have a lower energy (by about 50 meV) than the $d_{xz}/d_{yz}$ orbital \cite{joshua2012universal}. Thus, beyond a certain value of the sheet charge carrier density $n^*$  the system can undergo a Lifshitz transition from a single band occupancy to a multiple band occupancy.  Across the Lifshitz transition, the electrical transport characteristics of the q-2DEG changes quite significantly: the Hall coefficient becomes a non-linear function of the magnetic field \cite{joshua2012universal,smink2017gate}, a novel transient superconducting state appears~\cite{daptary2017observation} and the longitudinal Hall resistance becomes anomalous \cite{joshua2013gate}. The value of $n^*$  has been experimentally found to lie in the range of $(1.7 - 2.9) \times$ 10$^{13}$cm$^{-2}$ \cite{joshua2012universal, eerkes2013modulation,hurand2015field,smink2017gate}. Evidence of  multiband transport has been seen in other experiments as well as, e.g., in optical transmission spectroscopy \cite{seo2009multiple} and thermal transport measurements~\cite{lerer2011low}. 

Resistance fluctuation measurements are a well established technique for studying charge carrier dynamics in condensed matter systems. In a previous publication we have reported the existence of large non-Gaussian resistance fluctuations around the superconducting transition temperature $T_C$ in LaAlO$_3$/SrTiO$_3$~\cite{daptary2016correlated}. These fluctuations were found to arise from the percolative transition of a Josephson-coupled superconducting networks. In this article, we concentrate on a temperature range slightly above the superconducting $T_C$. Previous studies have shown that at around $T \sim 1.8T_C$, superconductivity is a hidden order which manifested itself as a transient phase under a large perpendicular time varying magnetic field at charge carrier densities $n>n^*$~\cite{daptary2017observation}.  There have also been previous reports of the observation of pseudogap \cite{richter2013interface} over this temperature range implying the presence of superconducting fluctuations without global phase coherence.  These observations have been attributed to the fact that multiple bands might contribute to the transport in this system. This prompted us to study electrical transport and noise over this temperature regime and to look for possible signatures of effect of multiband transport. In this article, we  present results of the measurements of resistance fluctuations (noise) at $T=1.8T_C$ as the chemical potential is tuned across the  Lifshitz transition. We observe a drastic change in the electrical transport properties as the system undergoes the transition. We attribute these changes to the difference between the nature of scattering of the charge carriers in the different bands.  

\begin{figure}[t]
	\begin{center}
		\includegraphics[width=0.5\textwidth]{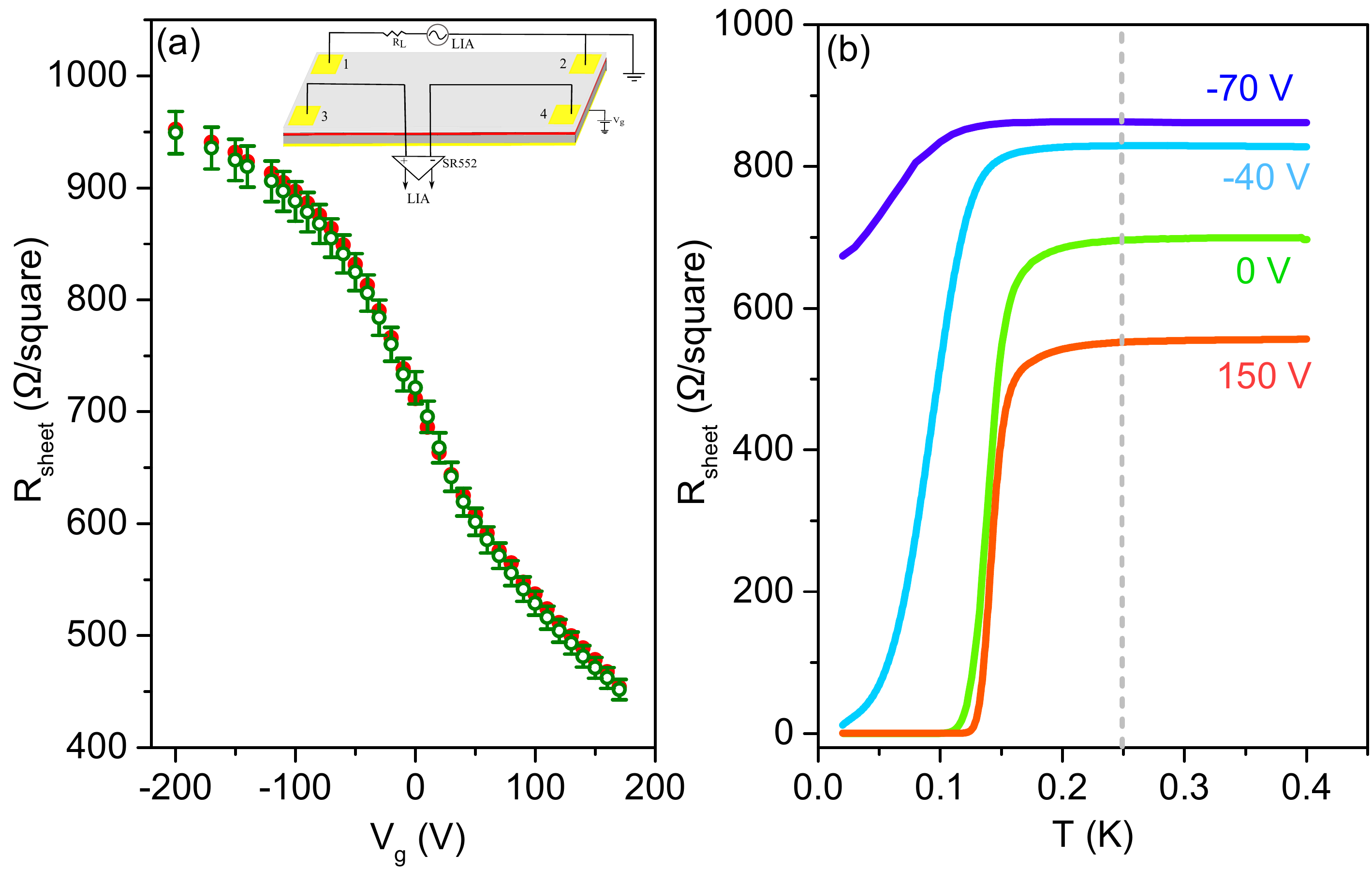}
		\small{\caption{(a) Plot of sheet resistance $R_{sheet}$ versus $V_g$  measured at $T$=245~mK and $B$=0~T (red filled circles). Shown also are the values of $R_{sheet}$ calculated using mobility and number density extracted from Hall measurements (see text for details). Inset: A schematic of the device structure. The red shaded area represents the q-2DEG located between LaAlO$_3$ and SrTiO$_3$ (001). For both resistance and noise measurements, an ac current is passed between the contacts marked 1 and 2 and the voltage drop between the contacts marked 3 and 4 is amplified by a low noise preamplifier (SR552) and detected by a lock-in amplifier (SR830). A thin layer of Au evaporated on back side of the SrTiO$_3$ acted as the gate electrode (b) Plots of the measured sheet resistance $R_{sheet}$ as a function of temperature at different $V_g$. The grey dotted line marks the temperature at which the noise data reported in this article were taken. \label{fig:figure1}}}
	\end{center}
\end{figure}

\begin{figure}[t]
	\begin{center}
		\includegraphics[width=0.5\textwidth]{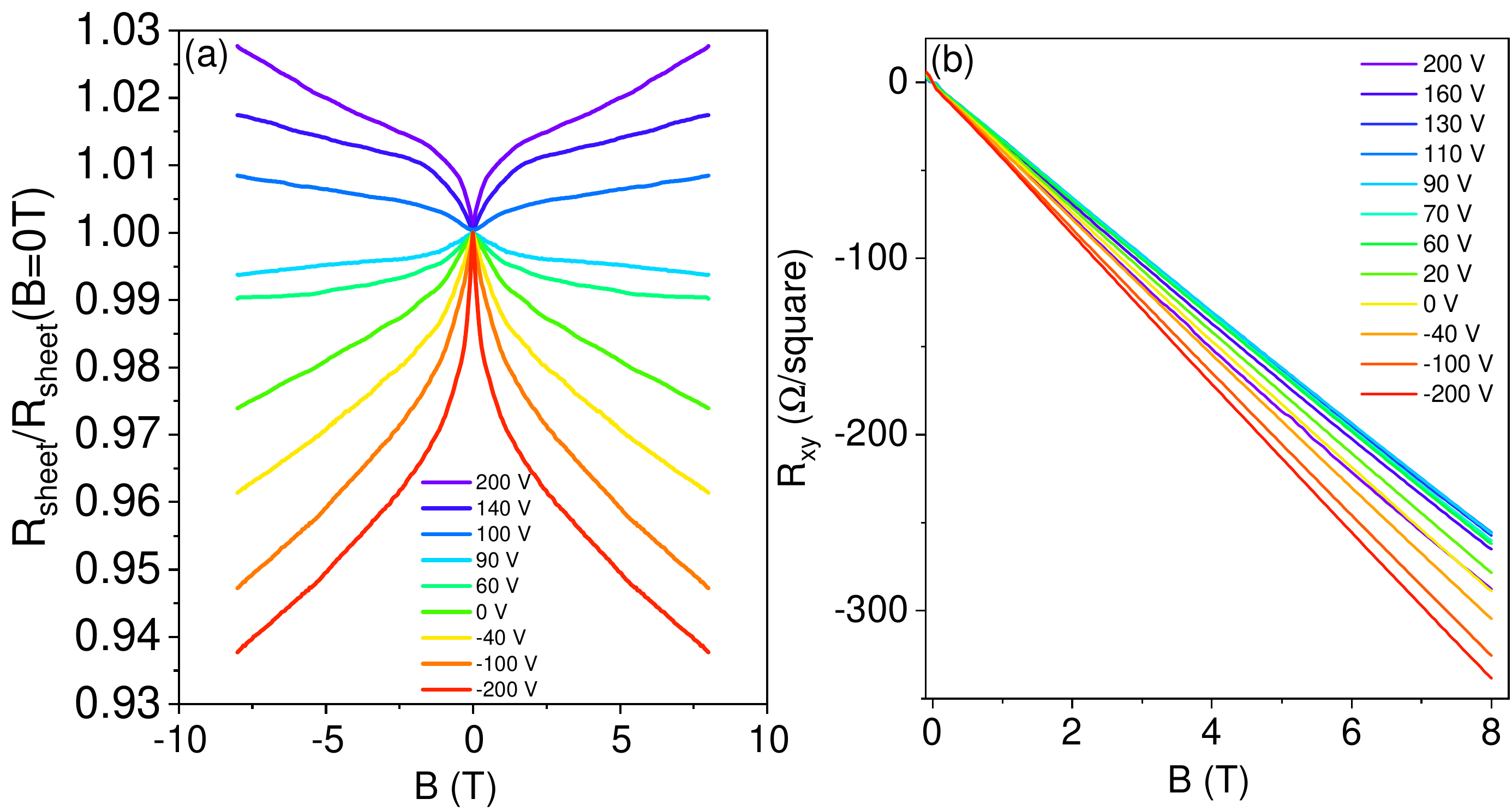}
		\small{\caption{(a) Longitudinal magnetoresistance (normalized by $R_{sheet}$(B=0))  and (b) Hall resistance as a function of magnetic field $B$ at different values of $V_g$. The measurements were taken at $T$=245~mK}. \label{fig:figure2}}
	\end{center}
\end{figure}
Our measurements were performed on samples with ten unit cells (u.c.) of LaAlO$_{3}$ grown by pulsed laser deposition on TiO$_{2}$ terminated (001) SrTiO$_{3}$ single crystal substrates of thickness 0.5 mm (for details of  sample preparation see Ref. \cite{kumar2015enhanced}). The charge carrier density at the interface was controlled by a back-gate voltage $V_g$ with the SrTiO$_3$ acting as the dielectric material. Electrical contact pads were prepared on top of the LaAlO$_3$ substrate by thermal evaporation of 5~nm Cr followed by 70~nm of Au. These contact pads were ultrasonically wire bonded to the measurement chip-carrier  - a process which is known to breakdown the 10 uc. of LaAlO$_3$ and give Ohmic contacts to the underlying electron gas \cite{caviglia2008electric,joshua2012universal,shalom2010tuning}. The carrier density at the interface was modulated using back gate voltage $V_g$. A schematic of the device structure is shown in the inset of Fig. \ref{fig:figure1}(a) along with the electrical connections. The temperature dependence of the resistance was measured in a cryo-free dilution refrigerator equipped with a 16~T magnetic field and the noise measurements were performed at 245~mK in a He-3 refrigerator (base temperature of the refrigerator) with  a temperature stability of better than $\pm$~0.5 mK. Measurements were performed on two different devices. In this article we concentrate on the data from one particular device, and the data from the other device was qualitatively similar. 

\begin{figure}[t]
	\begin{center}
		\includegraphics[width=0.5\textwidth]{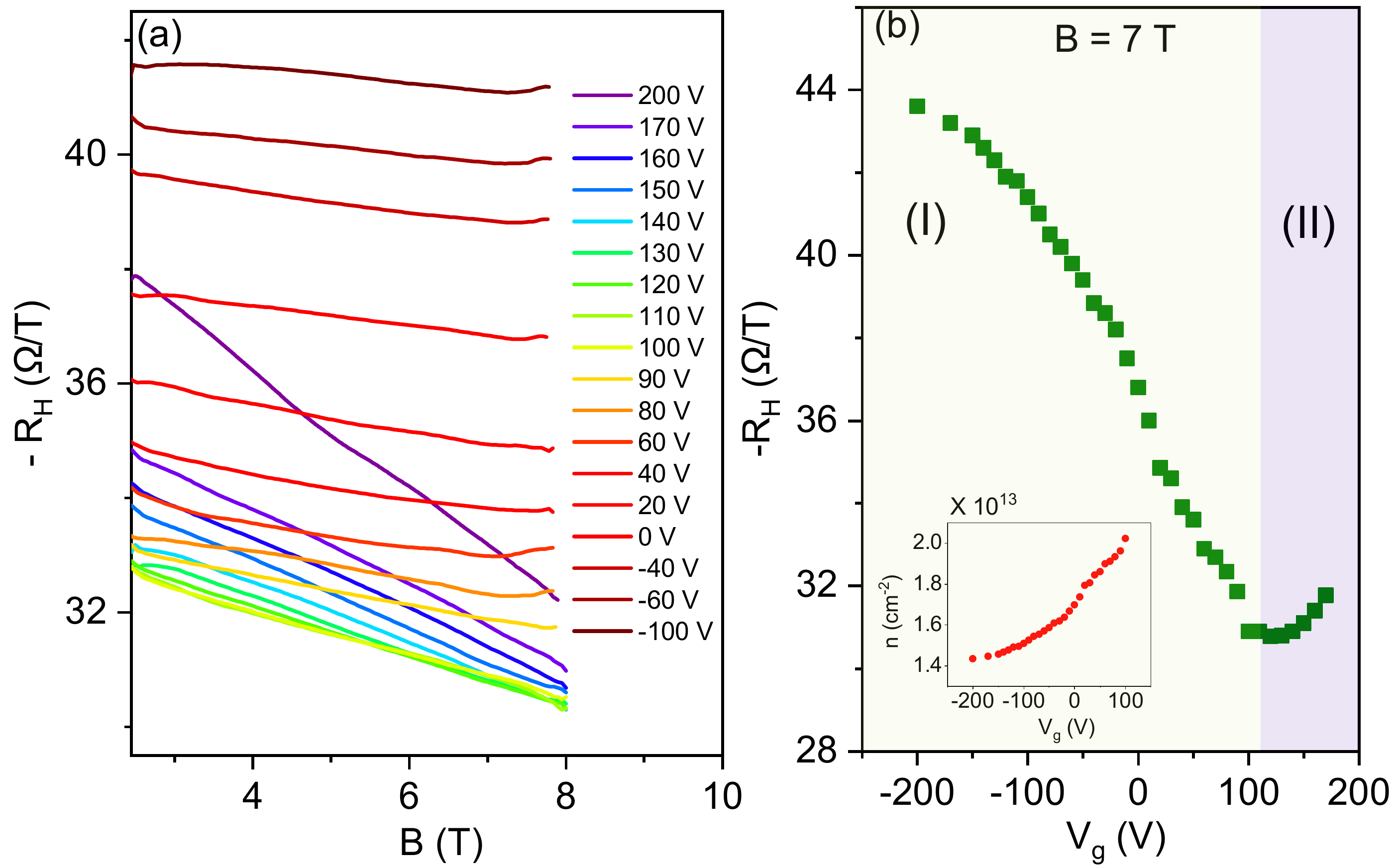}
		\small{\caption{ (a) Hall coefficient $R_H=(dR_{xy}/dB$) as a function of magnetic field at different values of $V_g$. (b) Hall coefficient $R_H$ at $B = 7$~T plotted as a function of $V_g$. The inset: $V_g$ dependence of the charge carrier density calculated from the Hall coefficient in the regime $V_g<V_g^*$.  Measurements were performed at 245~mK. \label{fig:figure3}}}
	\end{center}
\end{figure}

In Fig. \ref{fig:figure1}(a) we plot the sheet resistance $R_{sheet}$ as a function of back gate voltages $V_g$ at 245 mK. The data have been taken for decreasing gate voltages. We found that upon initial cool down, the resistance  $R$ versus $V_g$ curves were always hysteretic and non-reproducible. Repeated sweeps between the maximum and the minimum values of $V_g$ at the lowest temperature stabilized the resistance values and the $R-V_g$ curves became reproducible while still remaining hysteretic. This is a common feature of this system and has been reported by several groups \cite{biscaras2014limit,rossle2013electric}. We have taken care to always perform the noise and magnetoresistance measurements only for decreasing gate voltages, going only in one direction of $V_g$ sweep -- from +200 V to -200 V. The monotonic decrease in $R_{sheet}$ with increasing $V_g$ seen in the plot has been reported previously by several groups~\cite{caviglia2008electric,dikin2011coexistence,daptary2017observation} and  confirms that the charge carriers are electrons. In Fig. \ref{fig:figure1}(b) we plot the sheet resistance $R_{sheet}$ as a function of $T$ at few representative values of $V_g$. With changing $V_g$,  both the superconducting transition temperature $T_c$ and normal state resistance change. The subsequent measurements reported in this article were all  performed at $T$=245~mK ($T/T_{C0} \sim$ 1.8, $T_{C0}$ being the mean field superconducting transition temperature at $V_g$=0~V), marked by a vertical dashed line in Fig. \ref{fig:figure1}(b).

In Fig. \ref{fig:figure2}(b), we plot the Hall resistance at a few representative  values of $V_g$. We notice that above a characteristic value of  the gate voltage, which we denote as $V_g^*$ [$\sim$100~V for this particular device], both the Hall resistance $R_{xy}$ and the magnetoresistance $R_{xx}$ change in character. Below $V_g^*$, $R_{xy}$ is linear in magnetic field $B$, and its slope decreases with increasing $V_g$ signifying an increase in electron doping of the system. $R_{xx}$ over this field range shows a negative magnetoresistance (see Fig. \ref{fig:figure2}(a)). Above \vg,  $R_{xy}$ becomes non-linear in $B$ developing a distinct kink whereas $R_{xx}$ has a positive magnetoresistance. This change from a negative magnetoresistance to a positive magnetoresistance around a certain value of $V_g$ has been observed before in LaAlO$_3$/SrTiO$_3$ heterostructures and has been interpreted to be due to a transition from weak localization to weak anti-localization mediated by the large Rashba SOC present in this system \cite{caviglia2010tunable}. At low values of $V_g$ (single-band regime), quantum corrections to the Drude magnetoconductivity is ascribable to weak localization. At the single- to multiband crossover, the large enhancement in the density of states at the Fermi energy leads to an increase in the strength of the Rashba spin-orbit coupling. This causes the spin relaxation time to become much smaller than the inelastic scattering time leading to the appearance of weak anti-localization and consequently positive magnetoresistance.

In Fig. \ref{fig:figure3}(a) we plot the Hall coefficient $R_H$ (=$dR_{xy}/dB$) as a function of the magnetic field at different values of $V_g$. We find that for $V_g<100$~V, $|R_H|$ is almost independent of $B$ and decreases monotonically with an increase in $V_g$. For $V_g>100$ V, $|R_H|$ depends on magnetic field and increases with an increase in $V_g$. This value of $V_g=100$~V, where there is a quantitative change in the character of Hall resistance, we refer to as $V_g^*$. The non-linear Hall resistance observed for $V_g>V_g^*$ can have multiple possible origins  -- the ones most relevant to this system being Shubnikov-de Haas oscillations (SdH),  anomalous Hall effect (AHE) due to magnetization induced by the external magnetic field~\cite{shalom2010tuning,PhysRevB.80.180410} or the AHE due to conduction through multiple bands~\cite{PhysRevB.82.201407}. We do not observe SdH oscillations in our devices, presumably due to the low mobility of our samples~\cite{caviglia2010two}. The main signature of the AHE due to B-field induced magnetization is saturation of the  Hall resistance at high fields~\cite{lee2011phase}. The fact that we do not see such saturation at high fields rules out this scenario. We can also rule out the AHE due to magnetic moments by noting that the effect of ferromagnetism in this system is strongest at low doping levels whereas the observed non-linearity appears at high doping levels. This leaves us to consider the AHE induced by multiband transport as the  most plausible origin of the non-linear Hall observed by us~\cite{joshua2012universal,PhysRevLett.103.226802,PhysRevB.82.201407}. 

In Fig. \ref{fig:figure3}(b) we plot $|R_H(7T)|$ as a function of $V_g$. In the plot we define two regions: (I) $V_g<V_g^*$ (light yellow shaded region) where $|R_H(7T)|$ decreases monotonically with increasing $V_g$ and (II)  $V_g>V_g^*$ (light gray shaded region) where $|R_H(7T)|$ increases with increasing $V_g$. In region I ($V_g<V_g^*$) the hall resistance  is linear in $B$ and consequently the charge carrier density $n$ can be extracted using the form $n=-1/(R_H e)$. The value of $n$ (shown in the inset of Fig.~\ref{fig:figure3}(b)) increases monotonically with increasing $V_g$ over this range of gate voltages and matches well with that estimated by taking into account the dielectric constant of SrTiO$_3$~\cite{daptary2017observation}. These observations affirm that over this doping range, transport is dominated by a single-type of  carrier. Above \vg,  where the non-linearity in Hall resistance sets in, it becomes essential to consider the contributions of multiple-types of charge carriers.
\begin{figure}[t]
	\begin{center}
		\includegraphics[width=0.5\textwidth]{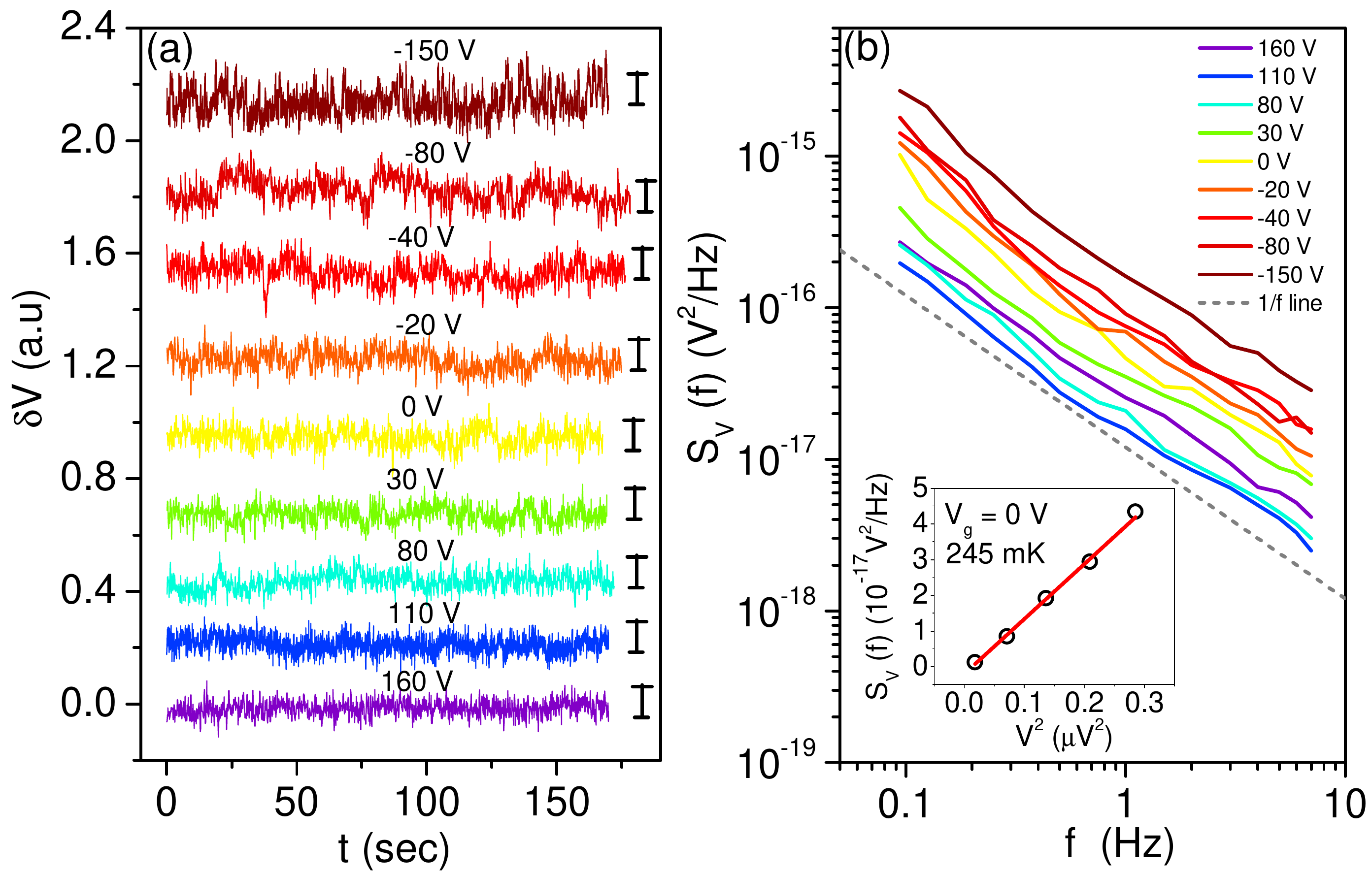}
		\small{\caption{(a) Time series of voltage fluctuations measured at different $V_g$'s at $T$=245~mK. The data have been vertically offset for clarity, and the vertical bar in each case represents 100~nV. (b)  Power spectral density of voltage fluctuations plotted as a function of frequency. The dashed line is a representative $1/f$ curve. The inset shows a plot of  $S_V(f=1~Hz)$ for $V_g$=0~V as a function of $V^2$. The red line is a linear fit to the data. \label{fig:figure4}}}
	\end{center}
\end{figure}

Given the fact that the q-2DEG is spread over multiple layers of Ti atoms, each presumably having different band configuration, it is difficult to develop a model that can describe quantitatively the observed Hall signal for $V_g>V_g^*$. Nevertheless, as a first approximation,  we analyze our Hall data using the two-carrier model following the formalism developed in Ref.~ \cite{joshua2012universal,bansal2012thickness,lerer2011low,PhysRevB.82.201407}:

\begin{align}
R_{xy}(B) = -\frac{B}{e}\frac{(n_1 \mu_1^2+n_2 \mu_2^2)+B \mu_1^2 \mu_2^2(n_1+n_2)}{(n_1 \mu_1+n_2\mu_2)^2 + B^2 \mu_1^2 \mu_2^2(n_1+n_2)^2}
\label{eqn:twoband}
\end{align}
with the constraint ($R(B=0T)=(en_1 \mu_1 + e n_2\mu_2)^{-1}$).  This model effectively considers the contribution from only one Ti layer or alternatively assumes that the electronic band structure and the band occupation in all the layers are the same. Here ($n_1$, $\mu_1$) are the carrier density and the mobility of $d_{xy}$ electrons whereas ($n_2$, $\mu_2$) are the corresponding quantities for the  $d_{xz}/d_{yz}$ electrons.  In Fig.~\ref{fig:figure1}(a) we show a plot of the $V_g$ dependence of $R(B=0) =(en_1\mu_1+en_2\mu_2)^{-1}$ calculated using the extracted values of $n_1$, $\mu_1$, $n_2$ and $\mu_2$. We find that the calculated resistance values match with the measured resistance to within $\pm$2\%. 

The critical carrier density $n^*$ (corresponding to \vg), at which the system undergoes a transition from a single-band-type to multiple-bands-type transport is $1.94 \times$ 10$^{13}$cm$^{-2}$ for this device - close to the number density range where the system is postulated to undergo a Lifsitz transition \cite{joshua2012universal,smink2017gate}. The Hall measurements thus indicate that, as discussed earlier, at low charge carrier densities, the itinerant  electrons occupy a single orbital. With increasing charge carrier density, the  system undergoes a Lifstitz transition and the higher energy  bands (predicted to be a mixture of bare $d_{xz}/d_{yz}$ bands modified by atomic spin-energy coupling~\cite{joshua2012universal}) begin to be occupied. 

To probe the dynamics of  charge carriers across this transition, we measured the resistance fluctuations in the sample at different $V_g$. The measurements were performed using a digital signal processing based ac technique~\cite{ghosh2004set} which allows simultaneous measurement of the background noise (Nyquist noise+instrumentation noise) as well as the bias-dependent noise originating due to resistance fluctuations of the sample. A low noise pre-amplifier (SR552) was used to couple the sample to a lock-in-amplifier (LIA). The carrier frequency $f_C$ of the LIA (we used $f_C$ = 228 Hz) was chosen lie in a region where the noise from the detection electronics was a minimum. 
The demodulated output signal of the LIA is digitized by a high-speed 16-bit analog-to-digital card and forms the time series of voltage fluctuations $\delta V(t)$. The voltage time series, consisting of 5 $\times$ 10$^6$ data points for each $V_g$, was digitally decimated and filtered to eliminate the 50~Hz line frequency. The power spectral density (PSD) of voltage fluctuations, $S_V(f)$ was obtained from this filtered time-series of voltage fluctuations by fast Fourier-transform over the spectral range of 90~mHz--8~Hz. The measurement system was calibrated down to spectral power $S_V(f) = 10^{-20}$ V$^2$/Hz by measuring the Johnson-Nyquist noise of a standard resistor. Typical time series of voltage fluctuations at several different values of $V_g$ are shown in Fig. \ref{fig:figure4}(a).  The corresponding $S_V(f)$ are plotted in Fig.~\ref{fig:figure4}(b). At all gate voltages, $S_V(f) \propto 1/f^\alpha$ with $\alpha \sim 0.9-1.1$.  $S_V(f)$  depended quadratically  on the voltage across the sample (see Fig. \ref{fig:figure4}(b) inset) indicating that the $1/f$ noise was generated from the resistance fluctuations of the sample. 

To compare the noise level of LaAlO$_3$/SrTiO$_3$ with other low-dimensional disordered superconductors, we calculated the Hooge parameter  defined as $\gamma_H = N\times \frac{f\times S_V(f)}{V^2}$; $N$ being the total number of charge carriers in the system. The value of $\gamma_H$  extracted at $V_g = 0$ V and $T$=245~mK is $\approx 2 \times 10^{2}$. This value is significantly lower than what is typically seen in other  low-dimensional disordered superconductors, e.g. NbN thin films ($ \gamma_H \approx10^{5}$)~\cite{koushik2013correlated}. 

\begin{figure}[t]
	\begin{center}
		\includegraphics[width=0.5\textwidth]{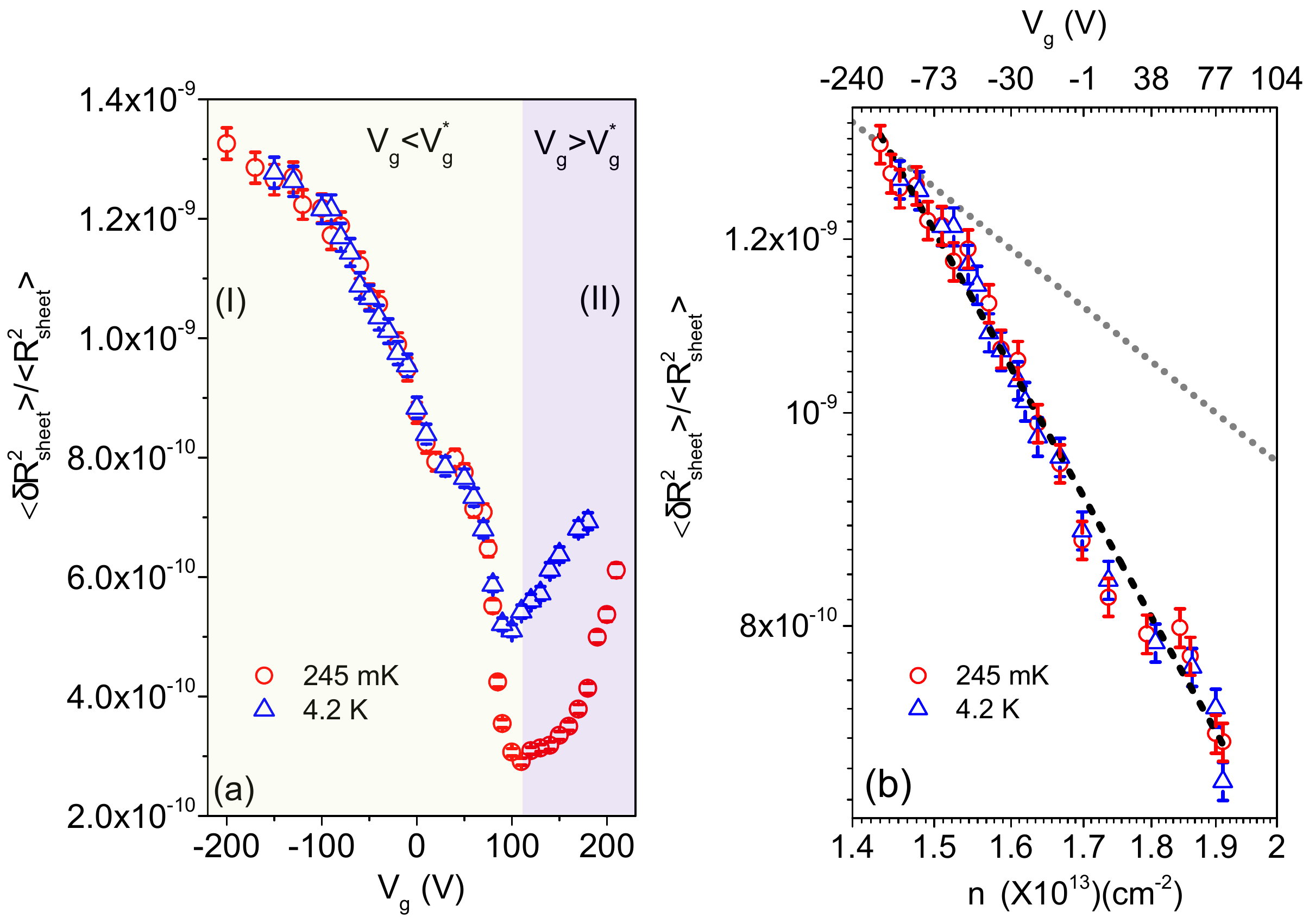}
		\small{\caption{ (a) Relative variance of resistance fluctuations $\mathcal{R} \equiv$\noise as a function of gate voltage $V_g$ at 245 mK (red open circles) and 4.2 K (blue open triangles). (b) Log-log plot of
				$\mathcal{R}$ versus the total charge carrier density $n$ extracted from Hall measurements at 245 mK and 4.2 K. The corresponding values of $V_g$ are marked on the top-axis. The black dashed line is a linear fit  to the data with slope $2.1\pm0.06$. The gray dotted line is a representative plot of slope -1.}.  \label{fig:figure5}}
	\end{center}
\end{figure}

A quantitative estimate of the dependence of the resistance fluctuations on $V_g$ can be obtained by evaluating the relative variance of resistance fluctuations  $\mathcal{R}$ defined as:
\begin{align}
\mathcal{R} \equiv \frac{<\delta R_{sheet}^2>}{<R_{sheet}^2>}=\frac{1}{V^2}\int_{90 mHz}^{8 Hz}S_V(f)df 
\end{align}
where $<\delta R_{sheet}^2>$ is the variance of the sheet resistance. In Fig. \ref{fig:figure5}(a), we plot $\mathcal{R}$ as a function of $V_g$ at 245 mK  and 4.2 K. In region I ($V_g<V_g^*$), the noise decreases monotonically with increasing $V_g$ until $V_g=V_g^*$.  On the other hand, for $V_g>V_g^*$, we find that the noise increased with increasing $V_g$. 

What can cause this dependence of the measured noise on $V_g$, or equivalently, on the charge carrier density? Resistance noise  in a system can arise either due to fluctuations in the mobility or in the charge-carrier number density:
\begin{align}
\mathcal{R}\equiv\frac{<\delta \mu^2>}{\mu^2}+\frac{<\delta n^2>}{n^2}
\end{align}

\noindent where $<\delta \mu^2>$ and $<\delta n^2>$ are variance in mobility $\mu$ and carrier density $n$ respectively. In a metallic system, the predominant contribution to noise comes from mobility fluctuations, in which case the noise scales as $\mathcal{R} \propto 1/n$ \cite{hooge19941}. In semiconductors, on the other hand the dominant source of noise is fluctuation in the charge carrier number density which leads to $\mathcal{R} \propto 1/n^2$~\cite{jayaraman19891}. In Fig. \ref{fig:figure5}(b), we show a plot of noise as a function of $n$ for $n<n^*$. We find that the data are well fitted with $\mathcal{R} \propto1/n^\alpha$, with $\alpha\approx 2.1\pm 0.06$ establishing that the resistance fluctuations over this range of doping is dominated by carrier density fluctuations. A physical mechanism of this process was proposed by McWhorter~\cite{jayaraman19891}, who showed that a major contribution to the resistance noise in semiconductors is from the  modulation of carrier density in the conducting channel due to trapping-detrapping from defect states. This requires the presence of an activated random process that can promote carriers from defect states in substrates or the bulk of the material to the conducting channel. The trap density extracted from the noise data using the McWhorter model~\cite{jayaraman19891} is $\approx 10^{13}$ cm$^{-2}$eV$^{-1}$ which matches well with prior reports of trap-densities measured in SrTiO$_3$ \cite{yadav2016amorphous}. A condition for the validity of the inverse square scaling of the noise with $n$ is that the charge carrier density  must be higher than the density of traps~\cite{vandamme2005additivity}, a condition that is satisfied in our system.   

The noise for $n>n^*$ has an upturn with increasing carrier density. Such an increase in noise with increase in $n$ is rarely seen. There have been reports of increase in resistance noise in superconducting systems as multiple bands become accessible to the system with varying temperature or bias~\cite{barone2011thermal,barone2014probing,barone2013electric}.  It has been proposed that a possible reason for this can be an increase in the available number of scattering channels at the transition from single-band to multiband transport.  In the absence of a theoretical model which quantitatively explains this behaviour, we leave a detailed explanation of the origin  of the excess noise in this regime to future endeavors. 

To summarize, we have studied in detail the carrier density dependence of resistance fluctuations in high quality LaAlO$_3$/SrTiO$_3$ heterostructures at temperatures slightly above the superconducting transition temperature. From magnetoreistance measurements we identify the critical number density $n^*$ at which the system undergoes  Lifshitz transition from a single-band to a multiband transport. For $n<n^*$, the measured noise arises due to fluctuations in the charge carrier density in the 2DEG. For $n>n^*$, we propose that the observed increase in noise with carrier density could be a manifestation of scattering of carriers between different available transport channels. Our results  emphasize the importance of inter-band scattering processes in systems which support carriers in multiple bands.

The authors thank R. C. Budhani for providing the samples. A.B. acknowledges funding from SERB, DST, Government of India.


%

\end{document}